# Regulation, Power, and the Compliance Paradox: A Longitudinal Study of Smart Homes

**Wael Albayaydh***
*ORCID:* 0000-0002-8334-5738
wael.albayaydh@cs.ox.ac.uk

**Ivan Flechais**
*ORCID:* 0000-0002-3620-0843
ivan.flechais@cs.ox.ac.uk

**Rui Zhao**
*ORCID:* 0000-0003-2993-2023
rui.zhao@cs.ox.ac.uk

Department of Computer Science, University of Oxford, Oxford, United Kingdom

**Corresponding author*

## ABSTRACT

Smart home technologies are increasingly embedded in domestic environments, yet their implications for power, privacy, and inequality remain unevenly understood—particularly in non-Western contexts. This paper presents a longitudinal, socio-technical study of smart home adoption in Jordan, examining how cultural norms, regulatory frameworks, and technological practices interact to shape domestic power dynamics.

Building on a prior 2022 study, we employ a two-phase grounded theory approach combining (1) a secondary analysis of 30 interviews and (2) 28 new interviews conducted in 2025 with returning and new participants across households, domestic workers, policymakers, and activists. This design enables a rare longitudinal perspective on how regulatory and technological changes reshape everyday surveillance practices.

Our findings reveal a compliance paradox: while Jordan's 2023 Data Protection Law increases awareness of privacy, it simultaneously enables new forms of exploitation by leaving domestic contexts unregulated. We identify three key dynamics: (1) the normalization of passive surveillance framed as convenience and legality, (2) regulatory ambiguity that redistributes responsibility into households, reinforcing existing hierarchies, and (3) constrained resistance among domestic workers navigating intensified monitoring.

We contribute a longitudinal account of how regulation does not simply fail but is appropriated within socio-cultural systems in ways that reproduce inequality. We conclude by outlining four intervention spaces spanning design, policy, and advocacy, highlighting the need for contextually grounded approaches to privacy and governance in smart home ecosystems.

## 1. INTRODUCTION

Smart home technologies are often framed as tools of convenience, efficiency, and modern living. However, their integration into domestic environments also reconfigures power relations, particularly in contexts where social hierarchies, cultural norms, and legal protections are unevenly distributed. These dynamics are especially pronounced in non-Western settings, where domestic labour, gender roles, and household authority structures intersect with emerging digital infrastructures.

This paper presents a longitudinal study of smart home adoption in Jordan, examining how power, privacy, and regulation evolve over time within domestic spaces. Building on a prior study, which analyzed 30 interviews conducted in 2022, we revisit the same socio-technical environment following the introduction of Jordan's 2023 Data Protection Law. Through a secondary analysis of the original dataset and 28 new interviews conducted in 2025—including 16 returning participants—we provide a rare longitudinal account of how regulatory and technological changes reshape everyday surveillance practices.

Prior work leaves three important gaps. First, little is known about how smart home surveillance practices and their justifications have evolved following Jordan's 2023 Data Protection Law. Second, limited attention has been paid to how AI-driven and ambient data collection interact with cultural norms and household power hierarchies to produce new privacy vulnerabilities, particularly for domestic workers. Third, prior studies have not examined whether earlier privacy and policy solutions remain valid under changing regulatory and technological conditions.

To address these gaps, this study investigates three research questions:

- RQ1: How have smart home surveillance practices and their justifications evolved following Jordan's 2023 Data Protection Law?
- RQ2: How do AI-driven and ambient data collection practices interact with cultural norms and household power hierarchies to shape privacy vulnerabilities, particularly for domestic workers?
- RQ3: To what extent do prior privacy and policy solutions remain valid under changing regulatory and technological conditions?

Guided by these questions, our analysis reveals what we term the **compliance paradox**: rather than mitigating harms, the introduction of data protection regulation increases awareness of privacy while simultaneously enabling new forms of exploitation. In practice, the law's ambiguity regarding domestic environments allows households to reinterpret surveillance practices as legitimate, shifting justifications from 'security' to 'legality.' This transformation redistributes responsibility for governance into private households, where existing socio-cultural hierarchies—particularly those affecting domestic workers—remain largely unchallenged.

Across our data, we observe three key shifts. First, surveillance becomes normalized through narratives of convenience and technological necessity, often obscuring its impact on vulnerable individuals. Second, regulatory ambiguity enables households to exploit legal gaps, reinforcing asymmetrical power relations. Third, domestic workers—particularly women—adapt to intensified monitoring through constrained forms of resistance, shaped by economic dependence and social risk.

### Contributions

This paper makes three primary contributions:

(1) A longitudinal empirical account of how smart home surveillance practices evolve over time in response to regulatory and technological change in a non-Western context.
(2) The conceptualization of the compliance paradox, demonstrating how regulation can be appropriated within socio-technical systems to reproduce, rather than reduce, inequality.
(3) A socio-technical perspective on governance, showing how responsibility for privacy and regulation is redistributed into domestic environments, where it is mediated by cultural norms and power hierarchies.

By foregrounding these dynamics, we extend existing work in HCI on privacy, power, and governance, highlighting the need for contextually grounded approaches that account for the interplay between law, culture, and technology in shaping everyday digital practices. The remainder of this paper presents background on smart homes, privacy, and regulation, followed by our methodology, findings, and implications for design and policy.

## 2. BACKGROUND

## 2.1 Context Overview and Smart Home Adoption in Jordan

Jordan, a Southwest Asian Arab country with a Muslim-majority population, is shaped by religious values, cultural norms, and socio-economic structures [35, 46, 77]. These factors influence family organization, gender roles, and authority within households [41, 62]. In such contexts, patriarchal structures often centralize decision-making power, which in turn shapes how technologies are adopted and used.

Smart home technologies—including surveillance cameras, smart speakers, and automated systems—have seen increasing adoption in Jordan in recent years[1]. While these technologies are often framed as tools of convenience and modernization, their integration into domestic environments intersects with existing social hierarchies.

Domestic labour in Jordan is largely performed by migrant workers, primarily women from Southeast Asia and Africa, who frequently experience constrained autonomy, limited legal protections, and restricted mobility [48, 84]. The Kafala (Sponsorship) system reinforces employer control over workers' residency and employment conditions, contributing to entrenched power asymmetries. Prior studies highlight how these inequalities manifest in everyday practices, including monitoring and mobility restrictions [5].

Despite increasing adoption, research on smart home technologies in Jordan and the wider MENA region remains limited. Existing studies suggest that users often prioritize convenience over privacy [6, 7], while awareness of privacy risks remains relatively low [65, 79]. These dynamics create conditions in which smart technologies may reinforce, rather than disrupt, existing socio-cultural hierarchies.

### *2.1.1 Domestic Workers in Jordan*

The International Labour Organization[2] and Jordanian Regulation No. 90/2009 [47] define domestic workers as individuals performing household maintenance tasks such as cleaning, cooking, laundry, child supervision, and related services. This category includes childcare providers, home-care aides, and property maintenance staff. The regulation [47] outlines reciprocal responsibilities between households and workers regarding privacy and safety.

Tamkeen's 2019 report [22, 84] estimates around 70,000 migrant domestic workers in Jordan, of whom only 54,000 are legally registered. Undocumented workers remain at high risk of detention and deportation. Most originate from Southeast Asia (Philippines, Indonesia, Bangladesh) and Africa (Ethiopia, Ghana). Jordan's labour framework [48] enforces a tiered wage system: 260 JOD for nationals, 245 JOD for general migrant labour, and 125 JOD for domestic workers, despite mandated food and housing. The kafala system grants employers extensive control over workers' residency, mobility, and documentation, often resulting in passport confiscation and restricted movement. Domestic workers cannot change employers, resign, or leave the country without written employer approval. Tamkeen received 328 complaints in 2019 [22], reporting wage withholding, passport retention, and denial of leave rights. These structural conditions make domestic workers especially vulnerable to surveillance and control within smart home environments.

## 2.2 Privacy and Cross-Cultural Perspectives

Privacy is widely recognized as a fundamental human right, but its meaning and interpretation vary across cultural, legal, and technological contexts [80, 91]. Classical definitions emphasize control over personal information and freedom from intrusion [76, 90], while contemporary perspectives highlight the role of social context and data practices in shaping privacy expectations [8, 71]. In IT contexts, privacy is often described as the ability to manage access to personal data stored and processed through digital systems [18].

Cross-cultural research demonstrates that privacy attitudes are influenced by cultural dimensions such as hierarchy, collectivism, and uncertainty avoidance [44, 61]. These factors affect how individuals perceive data sharing, surveillance, and acceptable boundaries of information flow [55]. As a result, privacy frameworks developed in Western contexts may not fully capture the realities of non-Western environments [92].

[1] Growth of smart devices in Jordan, Jordan Digital Strategy

[2] ILO-International Labour Organization

In Muslim and non-Western contexts, privacy is further shaped by religious values, social norms, and shared living practices. Prior work suggests that Islamic perspectives recognize privacy as a fundamental right, yet empirical research examining how these values interact with smart technologies remains limited [42, 72]. This gap highlights the need for contextually grounded studies that account for cultural and religious influences on privacy practices.

## 2.3 Smart Home Privacy and Bystanders

Smart home technologies introduce new privacy challenges by enabling continuous data collection within domestic spaces. These devices may capture information about both primary users and individuals who do not directly control them, including family members, guests, and domestic workers [1, 4, 9, 21, 56, 93, 94, 97]. Users often express general concerns about privacy, yet these concerns are frequently outweighed by perceived benefits such as convenience and security [34, 69, 98].

The concept of bystanders has become central in understanding these dynamics. Following Yao et al [96], bystanders are individuals who are subject to data collection without ownership or meaningful control over the devices. Research shows that bystanders often lack awareness of device presence or functionality and have limited ability to manage their privacy [2, 63, 64, 66]. These challenges are particularly pronounced in domestic environments, where social and economic factors constrain individuals' ability to assert their preferences.

Domestic workers represent a critical subgroup of bystanders. Prior studies indicate that they are frequently exposed to surveillance technologies while lacking authority to influence their deployment [17, 96]. Existing work has primarily focused on Western contexts, with limited attention to how these dynamics unfold in non-Western settings [24, 80]. This gap is especially relevant in regions where domestic labour is embedded within broader socio-cultural hierarchies.

## 2.4 Power Dynamics in Smart Home Environments

Smart home technologies are embedded within existing power structures that shape their use and impact. Differences in knowledge, authority, and socio-economic status influence who controls devices and how data is collected and used [12, 16, 31, 40, 73]. Research shows that individuals with greater authority within households often manage devices on behalf of others, enabling them to monitor or influence behaviour [32, 33, 70].

In domestic labour contexts, these dynamics may reinforce employer control, extending surveillance into workers' daily activities [17]. Even when workers are aware of monitoring, they may feel unable to object due to economic dependence or fear of repercussions [49]. Similar patterns are observed in workplace surveillance, where individuals trade privacy for employment stability [13, 60, 81].

Cultural norms play a central role in shaping these dynamics. Expectations around authority, obedience, and social roles influence how surveillance is perceived and justified [59]. As a result, smart home technologies often reinforce existing hierarchies rather than challenging them.

## 2.5 Regulation and Data Protection

Regulatory frameworks are critical in shaping how privacy is defined and enforced. Globally, data protection laws such as the GDPR have introduced principles of consent, transparency, and user rights, influencing regulatory approaches worldwide [15, 45, 75]. However, many of these frameworks do not explicitly address smart home environments, leaving gaps in how domestic data collection is governed [11, 14, 25, 74, 78, 86, 88].

Our review of data protection regulation in major jurisdictions highlights the broad spread of such frameworks, but also shows that smart homes remain largely unaddressed as a specific domain (Table 1).

**Table 1. Status of Data Protection Regulation in World Leading Countries**

| Country | Data Law | Regulatory Body | Year | Smart Home |
|---|---|---|---|---|
| Turkey | DPL[3] | KVKK | 2016 | No |
| Malaysia | PDPA[4] | Malaysia-PDP | 2011 | No |
| Europe | GDPR | EDPS | 2018 | No |
| USA | State Dependent | State Dependent | 2022 | No |
| China | PIPL[5] | CAC | 2021 | No |
| Brazil | LGPD[6] | ANPD | 2021 | No |
| India | DPDP IT Act[7] | India DPA | 2023 | No |
| Canada | PIPEDA[8] | OPC | 2000 | No |

Jordan's Personal Data Protection Law, introduced in 2023 [52], represents a significant development in formalizing data protection rights. While the law draws on international models and introduces mechanisms such as consent and data processing controls, it does not clearly define how these protections apply within private domestic spaces or to smart home technologies.

Existing legal frameworks in Jordan address aspects of surveillance and data protection, including constitutional provisions and sector-specific laws (Table 2, Figure 1). However, these frameworks remain fragmented and limited in scope, particularly in relation to domestic environments. This ambiguity creates uncertainty regarding rights and responsibilities, especially for individuals who are not device owners.

**Table 2. Existing Laws in Jordan**

| Law | Articles | Coverage | Smart Home |
|---|---|---|---|
| Personal Data Protection Law[9] | All | Personal Data | Limited |
| Cyber-Crime Law[10] | 11,13 | Surveillance, Defamation | Limited |
| Telecommunications Law[11] | 56 | Communications | No |
| Penal Code[12] | 348 | Private Data | Limited |
| Labour Law[13] | All | Work Conditions | No |
| Access to Info Law[14] | 13 | Gov Data | No |
| Credit Info Law[15] | 8 | Credit Data | No |

**Fig. 1. Timeline: Smart Technology Adoption vs Privacy Regulation in the MAME Region**

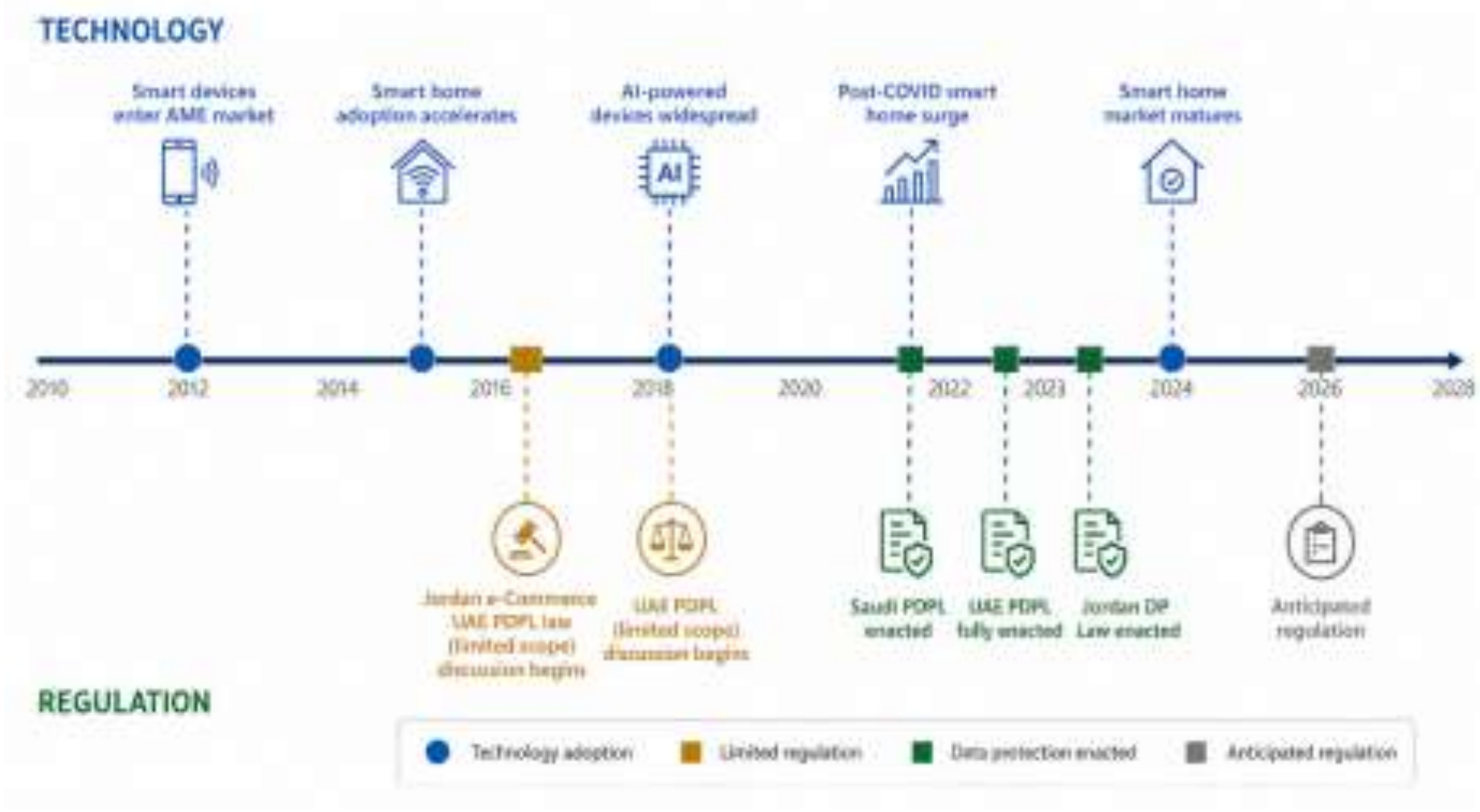

## 2.6 Summary and Research Gap

Prior work highlights three key limitations. First, most research on smart home privacy and surveillance has been conducted in Western contexts, limiting understanding of how these technologies operate in different socio-cultural environments. Second, while existing studies identify power asymmetries and bystander vulnerabilities, they often provide only static snapshots rather than examining how these dynamics evolve over time. Third, the role of regulation in shaping everyday surveillance practices remains underexplored, particularly in domestic settings.

This study addresses these gaps through a longitudinal analysis of smart home adoption in Jordan, examining how cultural norms, technological practices, and regulatory frameworks interact to shape power and privacy over time.

# 3. METHODOLOGY

This paper employs a mixed-methods approach to analyze the evolution of power dynamics within Jordan's smart home ecosystem, building on a prior research from 2022. By combining a secondary analysis of the original 2022 dataset with new qualitative interviews from 2025, the study enables a longitudinal comparison. This approach provides insights into how Jordan's 2023 Data Protection Law has reshaped surveillance practices and power relations within domestic spaces, allowing the framework to track shifts in data collection, regulatory impacts, and cultural rationales over three years.

The methodology is anchored in grounded theory, supporting systematic analysis across two phases. The first phase involved a secondary analysis of the 2022 interview transcripts (n=30), reapplying the coding framework while incorporating new analytical perspectives to surface latent patterns. The second phase included semi-structured interviews in 2025 with 16 participants from the original cohort—households, domestic workers, policymakers, labor law experts, and activists—plus 12 new stakeholders. This sampling ensures continuity while capturing evolving attitudes and practices. Together, the dual-phase design corroborates original findings and highlights developments in the deployment and justification of smart technologies within Jordan's socio-cultural and regulatory landscape.

Participants were selected to capture diverse perspectives while maintaining methodological rigor. Returning participants with insightful prior accounts were purposively recruited, along with new participants representing key stakeholder groups. Data collection employed semi-structured interviews to balance cross-case comparability with flexibility to probe emergent themes. This design provides a robust empirical foundation to analyze the interplay between smart home technologies, cultural norms, and regulatory frameworks in shaping power relations, as discussed in Section §6. For clarity, this study uses the following definitions: Households are families using smart devices and employing domestic workers; Workers are domestic workers; Regulators are ICT policymakers in Jordan's public and private sectors; Labor Law Experts are labor law policymakers; and Activists are human and civil rights advocates. Participants are labeled by role and number: households [H01–H07], domestic workers [W01–W07], regulators [R01–R05], labor law experts [LE01–LE04], and activists [A01–A05]. Numbers distinguish participants within the same role (e.g., [H01] for Household 1). Returning participants retain original identifiers (e.g., [H01]), while new participants receive sequential numbers (e.g., [H13]).

## 3.1 Secondary Analysis

Secondary analysis constitutes a systematic methodological approach, characterized by defined procedural and evaluative stages, which employs pre-existing data to investigate new research questions distinct from those of the original inquiry [38, 43]. This study adhered to the established protocol delineated by Johnston [50], a process which involved formulating specific research questions, subsequently identifying and critically evaluating pertinent datasets, and culminating in their rigorous analysis.

### *3.1.1 Developing the Research Questions*

The first step in secondary analysis involves formulating research questions. A gap was identified in the longitudinal study of users' privacy in Jordanian smart homes. Prior studies examined privacy asymmetries and surveillance practices, but research remains limited in three ways.

First, earlier work captured power relations and data practices before Jordan's 2023 Data Protection Law, leaving unclear how new legal frameworks have reshaped domestic surveillance norms over time. This gap informed RQ1, which examines how smart home surveillance practices and their justifications have evolved following regulatory change.

Second, prior research identified ethical and cultural tensions but did not systematically address vulnerabilities linked to the normalization of passive or ambient data collection. AI-driven monitoring introduces subtle and pervasive oversight, especially in households where cultural norms and gender hierarchies already define asymmetric power. This gap informed RQ2, which examines how AI-driven monitoring interacts with household power relations to shape privacy vulnerabilities, particularly for domestic workers.

Third, no study has assessed the continued validity of prior privacy and policy solutions under these changes. Recommendations made in a pre-regulatory context may no longer be fully relevant after the 2023 Data Protection Law and new technological affordances. This gap informed RQ3, which examines whether prior privacy and policy solutions remain valid under changing regulatory and technological conditions.

Overall, these gaps reveal the lack of a comprehensive, longitudinal, and contextually grounded analysis of how data protection reforms, cultural norms, and smart technologies interact to reproduce or transform power imbalances in Jordanian households. This study addresses that gap by combining secondary analysis of the 2022 dataset with new 2025 empirical data, providing a rare longitudinal perspective on the evolution of surveillance, privacy, and regulation in a Muslim Arab Middle Eastern context.

### *3.1.2 Identifying the Dataset*

The selection of a dataset for secondary analysis was guided by a systematic review of smart home research in the MAME region. This review did not identify other datasets directly addressing the power and privacy dynamics of smart home technologies in Jordanian households. Consequently, the dataset from a prior study was selected due to its unique relevance to the present inquiry, its methodological rigor, and its direct alignment with the research questions introduced in Section 1. The scholarly contribution of this original work has been recognized in prior research.

The alignment is pronounced, as both investigations center on the complex and deeply embedded power hierarchies within Jordanian households. The qualitative nature of the original data is particularly advantageous, as its rich contextual detail supports grounded re-interpretation to investigate emergent phenomena, including the normalization of passive and ambient data collection, evolving justifications for domestic surveillance, and the relationship between regulation and everyday monitoring practices.

The dataset is particularly appropriate for addressing the study's three research questions. First, because the original data was collected prior to Jordan's 2023 Data Protection Law, it provides a critical pre-regulatory baseline for examining how surveillance practices and their justifications have evolved over time (RQ1). Second, its detailed accounts of domestic relationships, cultural norms, and household authority structures support renewed analysis of how AI-driven and ambient data collection interact with power asymmetries to shape privacy vulnerabilities, particularly for domestic workers (RQ2). Third, because the original study generated privacy and policy recommendations in a pre-regulatory context, the dataset provides a foundation for evaluating their continued validity under changing legal and technological conditions (RQ3).

Furthermore, the foundational approach—observing cultural practices in situ—is uniquely appropriate for analyzing the experiences of both technology adopters and marginalized subjects within domestic ecosystems, thereby enabling a nuanced understanding of how smart technologies both reflect and reinforce socio-technical inequalities [20].
Taken together, these characteristics make the dataset uniquely positioned to support the present longitudinal inquiry and to explore what emerges in our analysis as the compliance paradox: the tension through which legal reforms intended to strengthen privacy protections may, under particular socio-cultural conditions, be appropriated in ways that reproduce or intensify existing inequalities.

### *3.1.3 Evaluating the Dataset*

Prior to analysis, the primary dataset underwent a rigorous evaluation to ascertain its appropriateness for this study's objectives. Adopting the systematic, stepwise framework for secondary analysis proposed by Stewart and Kamins [82], we applied established evaluative criteria (e.g., [27, 29]) to assess the dataset's congruency, quality, and overall rigor.

Our methodical examination focused on four key dimensions: (a) the original study's research purpose and theoretical orientation; (b) the roles and expertise of the entities responsible for data collection; (c) the specific methods, timing, and context of data acquisition; and (d) the internal consistency and coherence of the collected information.

This meticulous pre-analysis confirmed the dataset's robustness and established its fitness for addressing the study's three research questions: examining changes in surveillance practices following Jordan's 2023 Data Protection Law (RQ1), analyzing how AI-driven and ambient data collection interact with household power asymmetries to shape privacy vulnerabilities (RQ2), and evaluating the continued validity of prior privacy and policy solutions under changing legal and technological conditions (RQ3).

## 3.2 Data Source

This study builds upon a prior investigation into smart home surveillance and power dynamics in Jordan. The original research adopted a qualitative approach to explore how households, domestic workers, policymakers, and activists understood and negotiated privacy within smart home contexts.

This study employs the same qualitative methodology to address the three research questions introduced in Section 1: examining how surveillance practices and their justifications have evolved following Jordan's 2023 Data Protection Law, analyzing how AI-driven and ambient data collection interact with household power asymmetries to shape privacy vulnerabilities, and evaluating the continued relevance of the privacy and policy solutions proposed in our original study under changing legal and technological conditions.

The data for this research consist of the original dataset of 30 interviews combined with a newly collected dataset of 28 semi-structured interviews. The interviews were conducted by Author#1, focusing on participants' lived experiences, emerging tensions, and interpretations of "appropriate" surveillance within smart home contexts.

By integrating both the 2022 and 2025 data collection waves, this study provides longitudinal insight into how smart technologies, evolving legal frameworks, and entrenched cultural values collectively influence privacy practices and domestic hierarchies in Jordanian households. This combined dataset also provides the empirical basis through which the compliance paradox emerged in our analysis, illuminating how regulatory reforms intended to strengthen privacy protections may be appropriated in ways that reproduce existing inequalities.

### *3.2.1 Recruitment*

The original study conducted a qualitative user study with 30 participants: 8 domestic workers, 7 household representatives, 7 policymakers, 5 human and civil rights activists, and 3 labor law experts from Jordan. For details on recruitment and demographics, see previous research in this area. To address this study's research questions, a new qualitative study was conducted with 28 participants: 16 returning from the original study and 12 newly recruited. This group included 7 domestic workers, 7 household representatives, 5 policymakers, 5 activists, and 4 labor law experts. The 12 new participants comprised 3 domestic workers, 3 households, 2 policymakers, 2 activists, and 2 labor law experts. Recruitment combined online advertising via Facebook and LinkedIn with snowball sampling to reach hard-to-access participants. Purposive and theoretical sampling ensured participants met inclusion criteria, with data collection continuing until thematic saturation. A screening questionnaire verified eligibility based on at least two years of role experience, awareness of smart devices and privacy issues, communication in English or Arabic, and willingness to participate in interviews and audio recordings.

We successfully reconnected with 16 participants from the original study and recruited 12 new participants. Recruitment of households and domestic workers was conducted through advertisements on social media, outreach via smart device vendors and domestic worker agencies, and snowball sampling. Regulators (from ICT and labor sectors) and activists were contacted through a range of private and public organizations, including ICT companies, mobile network operators, the Ministry of Digital Economy and Entrepreneurship[16], the Jordan Telecom Regulatory Commission (TRC[17]), Intaj[18], and JOSA[19]). The recruitment process faced challenges due to sensitivities surrounding data protection [54, 89], which were mitigated through snowball sampling techniques [10, 37]. This multi-pronged approach ultimately yielded a total of 27 new candidates: 7 households, 6 domestic workers, 5 ICT regulators, 4 labor law experts, and 5 activists. Participant familiarity with smart devices was assessed using Dreyfus' five-stage model of skill acquisition [30].

All 27 candidates were contacted via email and phone; 19 completed the screening questionnaire, and 12 new participants were ultimately recruited: 3 domestic workers, 3 households, 2 policymakers, 2 activists, and 2 labor law experts. Combined with retained participants, the final sample included 7 households, 7 domestic workers, 5 ICT policymakers, 4 labor law experts, and 5 activists, representing 14 households, 3 ICT companies, 3 government entities, 2 civil rights organizations, and 2 independent activists. To ensure ethical integrity, domestic workers whose employers were previous participants were excluded; this restriction did not apply to policymakers or activists. Interviews with households and domestic workers focused on socio-economic and power dynamics in Jordanian smart homes, domestic workers' agency, privacy perceptions, behavioral adaptations, and the impact of the 2023 Data Protection Law. Policymakers and activists discussed the perceived effectiveness of the new legislation in protecting privacy for both households and domestic workers. (For participant demographics, see Table-3, Table-4, and Table-5).

**Table 3. Demographic Information of Domestic Workers**

| P# | Participated in Original Study | Gender | Nationality | Age Group | Education | Job Type | | Competence with Smart Devices | Existing Smart Devices in Home |
|---|---|---|---|---|---|---|---|---|---|
| W01 | Yes | Female | Philippines | 30-39 | Diploma | Nurse | Full-Time | Proficient | Smart Camera, Smart Speaker |
| W02 | Yes | Male | Jordanian | 40-49 | BSc | Nurse | Part-Time | Novice | Smart Camera, Smart TV, Smart Light |
| W03 | Yes | Female | Philippines | 20-29 | High school | Baby Sitter | Full-Time | Novice | Baby Camera, Smart Heating System |
| W04 | Yes | Female | Bangladish | 30-39 | High school | Maid | Full-Time | Novice | Baby Camera , Smart Refrigerator |
| W05 | No | Male | indonesian | 20-29 | High school | Maid | Full-Time | Novice | Smart Camera, Smart Security System |
| W06 | No | Female | Philippines | 20-29 | High school | Nurse | Full-Time | Novice | Smart Camera, Smart Speaker |
| W07 | No | Female | Bangladish | 20-29 | High school | Maid | Full-Time | Novice | Baby Camera |

**Table 4. Demographic Information of Households**

| P# | Participated in Original Study | Gender | Nationality | Age Group | Degree | Job Type | Competence with Smart Devices | Used Smart Devices | Domestic Worker Info |
|---|---|---|---|---|---|---|---|---|---|
| H01 | Yes | Female | Bangladish | 40-49 | B.Sc. | Pharmacist | Expert | Smart Camera, Samsung Smart TV | Maid, Full-Time |
| H02 | Yes | Female | Jordanian | 40-49 | B.Sc. | No Job | Proficient | Google Home, Smart Camera | Baby Sitter, Part-Time |
| H03 | Yes | Male | Philippines | 30-39 | M.Sc. | Finance Manager | Expert | Smart Camera, Smart Door Lock | Maid, Full-Time |
| H04 | Yes | Female | Jordanian | 20-29 | M.Sc. | Mechanical Engineer | Expert | Merkury Smart Camera | Nurse, Part-Time |
| H05 | No | Male | Jordanian | 30-39 | B.Sc. | Doctor | Expert | Smart Camera | Nurse, Part-Time |
| H06 | No | Male | indonesian | 40-49 | B.Sc. | Developer | Proficient | Smart Camera, Smart TV | Maid, Full-Time |
| H07 | No | Female | Bangladish | 30-39 | M.Sc. | Teacher | Expert | Amazon Echo Dot | Maid, Full-Time |

**Table 5. Demographic Information of Policy Makers & Activists**

| P# | Participated in Original Study | Gender | Age Group | Degree | Domain/Field | Domain | Experience |
|---|---|---|---|---|---|---|---|
| R01 | Yes | Male | 40-49 | M.Sc. | Regulatory Expert | Private - ICT Sector | 12 years |
| R02 | Yes | Female | 30-39 | M.Sc. | Regulatory Expert | Private - ICT Sector | 8 years |
| R03 | Yes | Male | 40-49 | B.Sc. | Minister of ICT | Government - ICT | 12 years |
| R04 | No | Male | 40-49 | B.Sc. | Policymaker | Government - ICT | 4 years |
| R05 | No | Male | 40-49 | B.Sc. | Policymaker | Government - ICT | 7 years |
| A01 | Yes | Female | 40-49 | B.Sc. | Human & Civil Rights activist | Society Organization | 8 years |
| A02 | Yes | Female | 30-39 | M.Sc. | Human & Civil Rights activist | Society Organization | 5 years |
| A03 | Yes | Female | 30-39 | B.Sc. | Human & Civil Rights activist | Society Organization | 7 years |
| A04 | No | Male | 40-49 | B.Sc. | Human & Civil Rights activist | Society Organization | 4 years |
| A05 | No | Male | 30-39 | B.Sc. | Human & Civil Rights activist | Society Organization | 6 years |
| LE01 | Yes | Male | 40-49 | B.Sc. | Labour Law Experts | Ministry of Labour | 10 years |
| LE02 | Yes | Female | 30-39 | B.Sc. | Labour Law Experts | Lawyer | 8 years |
| LE03 | No | Female | 40-49 | M.Sc. | Labour Law Experts | Lawyer | 12 years |
| LE04 | No | Male | 40-49 | B.Sc. | Labour Law Experts | Lawyer | 8 years |

## 3.3 Methodology and Interviews

We developed three semi-structured interview guides, each tailored to a specific participant group: domestic workers, households, and policymakers and activists (including regulators, human and civil rights activists, and labor law experts). The interview scripts were structured following the funnel technique [23], beginning with broad, open-ended questions and gradually progressing to more specific and detailed inquiries. This approach facilitated rapport-building with interviewees and encouraged the elicitation of rich, nuanced information. By transitioning from general to focused questions, the interviewer was able to ensure comprehensive coverage of key themes without overwhelming participants at the outset.

Grounded Theory, following Strauss and Corbin's analytical framework [83], was employed for codebook development and theory generation, as it is particularly suitable for exploring under-researched domains. Grounded Theory enables the systematic construction of explanatory frameworks through iterative data collection, coding, and inductive reasoning [26]. This approach provided an in-depth understanding of the interplay between smart home power dynamics and data protection legislation in Jordan, supporting the formulation of context-sensitive insights and recommendations. By examining research questions from multiple stakeholder perspectives, Grounded Theory facilitated the identification of underlying perceptions, beliefs, and behavioral drivers shaping privacy practices and regulatory challenges.

All interviews were conducted remotely via Zoom and Facebook Messenger. With participants' oral consent, the interviews were audio-recorded. A trained researcher carried out the 28 interviews in both English and Arabic—24 in English and 6 in Arabic. The Arabic interviews were carefully translated by the researcher to ensure fidelity to participants' original meanings and viewpoints. The 28 semi-structured interview transcripts were analyzed using NVivo 12 Pro software.

To ensure the reliability and credibility of the codebook, we assessed inter-rater reliability [58] and employed triangulation consistent with grounded theory principles [51]. The empirical results of this study are presented in Section §5, followed by a discussion in Section §6, which provides an integrated analysis and offers recommendations for social, legal, business, and technical interventions aimed at addressing smart home power imbalances and enhancing privacy protection for both households and bystanders. Recruitment materials explicitly invited voluntary participation, and all participants joined the study voluntarily without financial compensation.

#### *3.3.1 Pilot Study*

We conducted three pilot interviews—one corresponding to each of the three semi-structured interview guides: households, domestic workers, and policymakers. The purpose of these pilot interviews was to assess the clarity, coherence, and appropriateness of the interview questions, as well as to identify any potential issues in the scripts prior to the main data collection phase. Following the pilot stage, no substantial revisions were required, as the interview guides were found to be clear and effective for their intended purposes.

#### *3.3.2 Interviews*

To address the study's research questions, the 2025 data collection engaged participants from five key stakeholder groups—households, domestic workers, ICT policymakers, labor law experts, and activists—to explore how smart home surveillance practices, privacy vulnerabilities, and regulatory understandings have evolved over time in Jordan.

Interview questions were designed to address the study's three research questions. To investigate changes in surveillance practices and their justifications following Jordan's 2023 Data Protection Law (RQ1), household participants were asked about the types of smart devices they used, the extent and purpose of their use, their awareness of data protection laws, and whether legal or normative understandings influenced their monitoring practices. To examine privacy vulnerabilities linked to AI-driven and ambient data collection under household power asymmetries (RQ2), domestic workers were asked about their interactions with employers, their experiences of monitoring, their privacy concerns and practices, and their capacity to assert or negotiate privacy preferences. To evaluate the continued relevance of prior privacy and policy solutions (RQ3), interviews with ICT policymakers, labor law experts, and activists focused on the adequacy of existing data protection frameworks in Jordan and potential strategies for mitigating harms associated with smart technologies.

To minimize response bias [28], interviews began with broad, open-ended questions addressing general privacy concerns without explicitly referencing power dynamics, marginalized groups, household authority structures, or privacy violations. This approach encouraged participants to share experiences organically, reduced priming effects, and allowed themes—including those later informing the compliance paradox—to emerge inductively through participants' own interpretations and accounts.

#### *3.3.3 Analyzing the Data*

The dataset consisted of interview transcripts from the original study together with new interviews conducted in 2025. All new interviews were audio recorded, transcribed verbatim, and analyzed using NVivo 12 Pro software. Author#1 translated the Arabic interviews with care to ensure participants' meanings were conveyed accurately without distortion.

Analysis was guided by the study's three research questions and conducted using iterative open coding informed by Strauss and Corbin's Grounded Theory approach [36, 83]. Both authors participated in data collection and analysis. Throughout coding, Author#2 consulted with Author#1 for clarification, interpretation, and contextual insight, while Author#1 annotated transcripts to provide additional cultural and regulatory context.

Coding focused on phenomena relevant to the research questions, including changes in surveillance practices and their justifications following Jordan's 2023 Data Protection Law (RQ1), privacy vulnerabilities associated with AI-driven and ambient data collection under household power asymmetries (RQ2), and tensions surrounding the continued validity of prior privacy and policy solutions (RQ3).

During initial open coding, 154 distinct codes were identified. Through constant comparison across both datasets, these codes were iteratively refined and expanded, with new codes introduced where warranted. Both authors then organized codes into broader themes through axial coding and overarching categories through selective coding, ultimately finalizing 166 codes systematically organized into the themes presented in Section 5.

To strengthen coding consistency and credibility, both authors cross-checked codes against selected transcripts, reviewed interpretations, and resolved differences through discussion to agree on a final codebook. Agreement across coders was assessed, yielding an average Cohen's kappa coefficient ($\kappa = 0.86$), indicating almost perfect agreement [67].

Data saturation—when additional data no longer yielded significant new insights—was confirmed separately across participant groups [26, 39]. Credibility was further reinforced through triangulation [51] by randomly selecting nine participants (three household representatives, three domestic workers, one regulator, one labor law expert, and one activist) to review and comment on the identified codes and themes. Participants confirmed the thematic structure, and their feedback contributed refinement to codes without generating new categories.

Finally, the authors examined how codes evolved longitudinally and clustered them into overarching themes reflecting shifts in surveillance, privacy, governance, and power relations under Jordan's emerging smart home ecosystem following the 2023 Data Protection Law. Through this process, the compliance paradox emerged inductively as an interpretive framework for understanding how legal reforms intended to strengthen privacy protections could be appropriated in ways that reproduce or intensify existing inequalities.

#### *3.3.4 Research Position and Ethics*

Our research focuses on power relationships and their influence on technology use. We reflect on our positionality [95] and its potential impact on both participants and research outcomes. The central question guiding our work examines how power dynamics shape technology use, particularly in relation to ethical principles such as freedom, fairness, and accountability.

To minimize bias, we implemented several safeguards: we designed the interview guide to avoid leading questions, framed the study as an exploration of privacy and interpersonal effects of smart technology, and employed two independent coders to analyze and cross-validate the data. We prioritized the protection of participants' anonymity and privacy while remaining attentive to any signs of unlawful treatment—which were not observed.

Given the exploratory nature of this study, we maintained a neutral stance, focusing on documentation rather than intervention when potentially unfair uses of technology were identified. This research domain poses inherent challenges, and future studies may uncover further cases of exploitation or unethical practices, emphasizing the need for standardized research protocols to identify, evaluate, and address such issues.

We acknowledge the limitations of this study, including possible biases influencing our findings, yet remain confident in their overall validity. Ethical approval for this research was granted by the Research Ethics Committee of our institution. Participants provided oral consent and were assured that their data would remain strictly confidential. Interview transcripts were encrypted and securely stored. Participants retained the right to withdraw at any time without explanation, with the assurance that their data would be excluded if they chose to withdraw. No participant chose to withdraw from the study.

# 4. LIMITATIONS

As with all qualitative research, this study is subject to several limitations that warrant consideration.
Selection of Participants: The participant sample primarily consisted of individuals who deliberately chose to adopt and engage with smart home technologies. Consequently, these intentional users do not fully represent the broader population of smart home users in Jordan. Non-users and bystanders may possess distinct perspectives, experiences, and concerns regarding such technologies. Moreover, individuals with heightened privacy or security apprehensions might deliberately avoid adopting smart devices altogether. Future research should therefore examine the views of various bystander groups (e.g., guests, passersby, or other categories of domestic workers) within Jordanian smart home environments.

**Language**: Participants were given the option to conduct interviews in either English or Arabic. While the majority of interviews were conducted in English, seven interviews were held in Arabic to accommodate native speakers. These Arabic interviews were meticulously translated into English to ensure the preservation of participants' original meanings and insights. Although some non-native English speakers occasionally faced difficulties articulating their thoughts, we believe this did not materially affect our analysis, as findings were cross-validated across all interviews.

**Recruitment**: Participant recruitment was particularly challenging due to legal constraints and the sensitivity surrounding issues of privacy and data protection. The delicate nature of these topics among regulators and key stakeholders may have shaped or constrained their responses. To mitigate such effects, participants were thoroughly briefed on the study's strict data protection procedures, including encryption protocols and compliance with the GDPR[21]. We also employed snowball sampling [37], a widely recognized method for reaching hard-to-access populations [10, 85].

**Topic Sensitivity**: Given the sensitive nature of privacy, surveillance, and regulatory issues, some participants may have offered cautious or incomplete responses to protect institutional interests or personal reputations. To minimize this bias, we clearly communicated our procedures for data anonymization, encryption, and GDPR compliance, which encouraged participants to respond openly.

**Research Quality**: The quality of qualitative research outcomes depends significantly on the researcher's skills and reflexivity, and may be influenced by personal bias. Inexperienced interviewers can face challenges in eliciting rich, nuanced data, which may lead to the omission of critical insights [19, 57]. To address this, the lead researcher received training in interview design and qualitative inquiry, employing neutral and open-ended questions to minimize bias, as detailed in Section §3.3.4.

**Bias**: Self-reporting bias represents an inherent limitation of interview-based studies [53]. Participants may forget events, misinterpret past experiences, or adjust their responses to align with perceived social desirability or researcher expectations [87]. To enhance validity and reduce such biases, we deliberately avoided leading questions and emphasized open-ended formats that encouraged participants to share experiences in their own terms. This approach enhanced both the comprehensiveness and credibility of our findings.

**Sample Size**: The qualitative nature of this research necessarily limits the breadth and diversity of the participant pool. Recruiting regulators willing to discuss sensitive topics such as privacy and data protection proved particularly difficult, further constraining the representativeness and heterogeneity of the sample.

**Generalization**: As this study employs a qualitative methodology, its findings are intended to provide in-depth, contextualized
understanding rather than statistical generalization. The insights reported here reflect the lived experiences and perspectives of our participants and should not be extrapolated to all populations. Any hypotheses or theoretical propositions derived from this grounded analysis warrant further empirical validation in subsequent studies to assess broader applicability.

**Research Validity**: Finally, we recognize the limitations associated with relying solely on Inter-Rater Reliability (IRR) to validate our analytical process. While IRR is a useful indicator of coder agreement, interpretation remains inherently subjective. High agreement may sometimes represent superficial consensus rather than deeper conceptual alignment. Moreover, IRR assesses coding consistency but does not capture broader aspects of validity—such as the contextual relevance of themes or the depth of theoretical interpretation—nor does it integrate direct participant validation, thereby leaving room for rater bias. Despite these limitations, IRR remains a valuable metric for promoting consistency and transparency. To strengthen credibility, it should be complemented by peer debriefing, multiple triangulation strategies (methodological, data, investigator, and theoretical), and iterative validation, particularly during early theme development and theory construction.

# 5. RESULTS AND FINDINGS

This section presents a longitudinal analysis of power dynamics, surveillance practices, and privacy experiences in Jordanian smart homes. Drawing on both the original 2022 study and the follow-up 2025 study, we examine how technological adoption, socio-cultural norms, and regulatory interventions interact over time to shape domestic data practices and lived experiences. Rather than treating smart home technologies as neutral tools, the findings demonstrate how they are embedded within and reinforce existing social hierarchies, particularly in relation to domestic labor and authority structures.

Across both datasets, a central phenomenon emerges: the compliance paradox. While the introduction of the 2023 Data Protection Law increases awareness of privacy and introduces formal legal discourse, it simultaneously enables new forms of surveillance and control. This occurs through regulatory ambiguity, selective interpretation, and the normalization of monitoring within everyday household routines. The paradox highlights that regulation does not necessarily reduce harm but can reshape how surveillance is justified, enacted, and experienced.

To provide a clear analytical structure, we first present the original study findings as a baseline, followed by the follow-up study findings, and then outline the longitudinal transformations observed between 2022 and 2025. We conclude with a synthesis that connects these findings to broader discussions of governance, power, and socio-technical systems.

## 5.1 Original Study Findings (2022): Foundations of Power and Surveillance

The original study establishes a baseline for understanding how smart home technologies intersect with socio-cultural and economic structures in Jordanian households. The findings were organized into two categories: (1) Smart Home Power Dynamics and (2) Perspectives on Mitigating Smart Home Power Dynamics (Table 6). These categories reveal that surveillance practices are not simply technical features but are shaped by broader systems of authority, dependency, and inequality that structure domestic life.

**Table 6. Summary of Categories and Themes of Original Study**

| Categories | Themes | Sub-Themes |
|---|---|---|
| Smart Home Power Dynamics | Contextual Power Dynamics | Smart Home Social Power Map |
| | | Household Privileges |
| | | Contextual Norms and Religious Background Influence Users' Relations and Privacy Considerations |
| | | Women Experience Reduced Power and Rights |
| | Economic Power Dynamics | Imbalanced Economic Power Dynamics |
| | | Workers Compromise Privacy Rights |
| | Smart Devices Impact Users Relations | Privacy Concerns |
| | | Strained Relationship |
| | | Smart Home Devices Reinforce Asymmetrical Power Dynamics |
| | Foreign Workers Suffer Contract Slavery | Domestic Foreign Workers Are Marginalized Group |
| | | Stringent Practices Towards Domestic Foreign Workers |
| Perspectives on Mitigating Smart Home Power Dynamics | Privacy Rights and Data Protection | Lack of Data Protection Regulation in Jordan |
| | | Economic-Contextual Influences on Privacy Rights |
| | Balancing Power Dynamics | Awareness is Important For Protection |
| | | Consider Household-Workers Needs |
| | | Regulation Modernization to Consider Smart Home Power Dynamics in Jordan |
| | | Aspirations For Innovative Solutions |

### *5.1.1 Contextual and Economic Power Structures*

The findings reveal that smart home technologies are embedded within deeply rooted socio-cultural and economic hierarchies that shape everyday interactions within the household. Household authority is influenced by cultural norms, gender expectations, and financial control, positioning employers as dominant actors who determine how technologies are deployed and used. Domestic workers, in contrast, occupy structurally vulnerable positions due to economic dependency and limited access to legal protections.

Participants described how surveillance technologies reinforce these hierarchies by making workers' actions visible and subject to continuous evaluation. One worker stated:

*"They watch everything I do... it feels like I have no space of my own."* [W03]

This reflects how surveillance extends beyond observation into psychological and spatial control. Workers anticipate monitoring and adjust their behaviour accordingly, demonstrating how power operates through both visibility and expectation.

#### *5.1.2 Smart Devices as Tools of Surveillance and Behavioural Regulation*

Smart home devices—including cameras, sensors, and connected applications—function not only as tools of observation but also as mechanisms for shaping behaviour. Households often justify the use of these technologies through narratives of safety and trust, framing surveillance as a protective measure. For example, one participant explained:

"*We need to make sure everything is safe when we are not at home*." [H02]

However, this framing contrasts with workers' experiences of surveillance as intrusive and restrictive. Workers reported modifying their behaviour, avoiding actions that could be misinterpreted, and limiting their movement within monitored spaces. In this sense, surveillance becomes a form of indirect control, where behaviour is shaped by the anticipation of observation rather than direct enforcement. This highlights how technology mediates authority in everyday contexts.

#### *5.1.3 Absence of Regulation and Informal Governance*

A defining characteristic of the 2022 context is the absence of formal regulatory frameworks governing smart home data practices. Privacy protections were largely informal and dependent on household norms rather than institutional oversight. Participants demonstrated limited awareness of data protection principles, and there were no mechanisms for consent, accountability, or redress. As one activist observed:

"*In practice, privacy is governed by whatever the household considered acceptable, not by any enforceable rights."* [A03]

A regulator similarly noted that there was no clear framework addressing how smart devices operated in domestic spaces, so oversight was effectively absent [R02]. This absence created a regulatory vacuum in which surveillance practices were normalized and rarely challenged. Governance operated through informal authority structures within the household, where decisions were shaped by cultural expectations and economic relationships.

### 5.2 Follow-Up Study Findings (2025): Regulation, Adaptation, and Transformation

The follow-up study examines how these dynamics evolve following the introduction of the 2023 Data Protection Law. While the law introduces formal recognition of privacy rights, its implementation reveals significant gaps between policy and practice. Participants described increased awareness of data protection concepts, yet emphasized that enforcement remains limited, particularly within domestic environments where oversight is minimal.

**Table 7. Summary of Categories and Themes of the Follow-Up Study**

| Categories | Themes | Sub-Themes |
|---|---|---|
| Evolving Data Practices and Regulatory Adaptation | Transformation of Data Collection Practices | Expansion of smart homes and smart devices |
| | | Data sharing with third-party service providers |
| | | Ambiguity in data ownership and consent |
| | Interpretations and Justifications of Data Use | Households justify monitoring as a form of protection and efficiency |
| | | Households' perception of surveillance as management accountability |
| | | Domestic workers' limited understanding of data purposes |
| | | Regulatory loopholes permitting over-collection |
| | Compliance and Enforcement under the 2023 Data Protection Law | Inconsistent awareness of legal responsibilities |
| | | Weak enforcement mechanisms and limited oversight |
| | | Privacy policies drafted without cultural contextualization |
| | | Lack of written or verbal consent practices |
| Emerging Vulnerabilities and Socio-Technological Power Shifts | New Socio-Digital Inequalities | Technological literacy gaps between households and workers |
| | | Economic constraints limiting data protection adoption |
| | | Continued marginalization of migrant workers despite new law |
| | Privacy and Security Vulnerabilities | Unclear accountability for privacy breaches |
| | | Workers' exposure to continuous audio–visual monitoring |
| | | Misuse of smart devices for behavioral profiling |
| | Reassessment of Prior Solutions and Future Directions | Need to re-evaluate ethical co-design principles for inclusivity |
| | | Potential for AI regulation alignment with worker rights |
| | | Integration of religious and social norms into privacy frameworks |
| | | Shift from surveillance-oriented design to empowerment-oriented design |

#### *5.2.1 Transformation of Data Collection Practices*

The follow-up study reveals a significant expansion in the scale and intensity of data collection within smart homes. Participants reported the integration of multiple interconnected devices, creating environments characterized by continuous monitoring. One participant noted:

> "*Every device has a sensor... it feels like monitoring never stops*." [W01]

Another domestic worker similarly explained that before there was maybe one camera, but now there are devices everywhere, and you do not know what is recording or when [W04]. This transition from episodic monitoring to persistent surveillance transforms the home into a data-rich environment. While households often perceive this expansion as beneficial for convenience, security, and oversight, workers experience it as a source of pressure and constraint.

#### *5.2.2 Data Ownership, Sharing, and Consent*

Participants expressed ongoing uncertainty regarding data ownership and sharing practices. Despite increased awareness of data protection laws, many users lack clarity about how data is collected, stored, and used. A household participant explained:

> *"You click agree, but you don't know where your data goes." [H02]*

As one regulator noted, consent becomes largely symbolic when individuals do not understand how data moves across devices, platforms, and third parties [R01]. For domestic workers, this uncertainty is even more pronounced. Workers reported limited understanding of data practices and rarely provided explicit consent:

> *"Nobody asked me... the camera was just there." [W07]*

#### *5.2.3 Legal and Managerial Justifications*

A notable shift in the follow-up study is the increasing use of legal language to justify surveillance practices within domestic environments. Unlike the 2022 findings, where monitoring was primarily framed in terms of safety and trust, participants in 2025 frequently referenced legal ambiguity as a basis for legitimizing their actions. For example, one participant stated:

> *"The law does not say we cannot use cameras." [H03]*

This reflects a broader interpretive strategy in which the absence of explicit prohibition is understood as implicit permission. This legal framing is further reinforced by managerial logics that position surveillance as a necessary tool for ensuring accountability and efficiency. As another participant explained:

> *"It's about making sure work is done properly." [H04]*

Together, these legal and managerial narratives reshape how surveillance is understood and enacted. Importantly, these justifications operate unevenly. While households can invoke legality and managerial necessity to legitimize surveillance, domestic workers lack the authority or knowledge to contest these interpretations.

#### *5.2.4 Weak Enforcement and Regulatory Gaps*

Despite the introduction of the 2023 Data Protection Law, participants consistently emphasized that enforcement mechanisms remain limited in practice, particularly within domestic environments. While the law establishes formal protections, its implementation is constrained by the absence of monitoring bodies, reporting mechanisms, and clear accountability structures. As one participant noted:

> *"There's a law, but no one checks." [LE01]*

The domestic context further complicates enforcement, as households are often treated as private spaces beyond the scope of formal oversight. As a result, compliance becomes selective rather than systematic. Households may invoke the law to justify surveillance practices—particularly when it supports their authority—while disregarding requirements related to consent, proportionality, or transparency.

Ultimately, weak enforcement does not simply limit the effectiveness of regulation but transforms its role. Rather than acting as a constraint on surveillance, the law becomes a flexible resource that can be interpreted and mobilized in ways that sustain and legitimize existing hierarchies within the home.

## 5.3 Longitudinal Transformations (2022-2025)

A comparative analysis of the two studies reveals several key transformations. In the 2022 dataset, surveillance was primarily justified through safety and trust. By 2025, this justification shifts toward legality, with participants increasingly relying on the absence of explicit legal restrictions to legitimize surveillance practices. As one household participant explained:

> *"If the law does not prohibit cameras inside my home, then I have the right to use them however I see fit." [H05]*

Monitoring practices also evolve from episodic and device-specific interactions to continuous and integrated systems. In addition, workers' responses change significantly—from discomfort and occasional resistance in the original study to adaptation and internalization of surveillance in the follow-up. Regulation itself shifts from absence to ambiguity, introducing legal frameworks that lack clear enforcement. Together, these transformations demonstrate that regulation reshapes rather than eliminates power dynamics, reinforcing the compliance paradox.

## 5.4 Emerging Vulnerabilities and Power Shifts

The follow-up study reveals the emergence of new vulnerabilities that arise from the interaction between technological expansion and regulatory ambiguity. While smart home adoption increases convenience and control for households, it simultaneously intensifies exposure and risk for domestic workers. These vulnerabilities are not solely technological but are shaped by socio-economic conditions, cultural norms, and unequal access to knowledge and resources.

Participants described how continuous monitoring alters the nature of domestic work, transforming it into a form of observed and evaluated labour. As one domestic worker explained:

> *"It feels like someone is always watching whether I am working fast enough, cleaning properly, even how long I pause. You start acting for the camera, not just doing your job." [W07]*

### *5.4.1 Adaptation and Internalization*

A key transformation observed in the follow-up study is the shift from resistance to adaptation among domestic workers. In the original dataset, workers expressed discomfort with surveillance and, in some cases, attempted to question or negotiate its presence. However, in the 2025 data, such resistance becomes less visible, replaced by forms of adaptation that reflect constrained agency.

One worker explained: "I cannot say anything." [W02], highlighting the limitations placed on workers' ability to challenge monitoring practices. Instead of resisting, workers described modifying their behaviour to align with perceived expectations. As one activist noted:

> *"What appears as acceptance is often survival. Workers adapt because refusal carries consequences they cannot afford." [A04]*

This shift represents a form of internalized compliance, where control is exercised not through direct enforcement but through self-regulation. Workers adjust their actions based on the anticipation of being observed, effectively embedding surveillance into their everyday routines.

### *5.4.2 Psychological Impacts*

The continuous presence of surveillance technologies has significant psychological and emotional effects on domestic workers. Participants described feelings of stress, anxiety, and discomfort associated with being constantly monitored. One worker noted:

> *"I feel like I can't relax." [W01]*

As one activist observed, the harm is not only that workers are watched, but that they begin living with constant psychological pressure, never fully at ease [A02]. A labour law expert similarly noted that continuous monitoring can produce emotional strain even without overt abuse, because the condition of permanent observation is itself a form of pressure [LE03]. These experiences indicate that surveillance extends beyond technical monitoring to influence emotional wellbeing.

#### *5.4.3 Socio-Digital Inequality*

The findings reveal that socio-digital inequalities play a critical role in shaping how individuals experience and respond to surveillance technologies. Differences in technological literacy, access to resources, and awareness of legal rights create uneven capacities for engaging with smart home systems. One participant stated:

> *"They know the apps... I just follow." [W06]*

As one regulator noted, rights on paper mean very little if the people most affected do not understand the technologies collecting data or the protections supposedly available to them [R05]. Economic factors further exacerbate these disparities, as access to privacy-enhancing technologies or legal resources is often limited.

#### *5.4.4 Behavioural Profiling*

The follow-up study indicates an increasing use of smart home technologies for behavioural profiling and evaluation. Surveillance systems are not only used to observe actions but also to interpret and assess behaviour over time. Participants described how data collected through cameras and sensors is used to form judgments about performance and reliability.

One worker explained: "They are judging everything I do." [W05], reflecting the perception that surveillance extends into evaluation. As one activist noted, what begins as monitoring can quickly become profiling, where ordinary actions are turned into judgments about trust, discipline, or worth [A01]. Behavioural profiling thus represents a significant expansion of surveillance, transforming it into a tool for ongoing assessment and reinforcing asymmetrical power relations.

### 5.5 Synthesis: The Compliance Paradox and Longitudinal Contribution

Our longitudinal analysis reveals that the introduction of formal regulation does not merely fail to eliminate surveillance-related harms; it actively reconfigures the moral and legal architecture used to justify them. We define this as the Compliance Paradox: a state where the maturation of a regulatory environment (2022-2025) provides a new vocabulary for the expansion of domestic monitoring rather than its restriction.

The strength of this longitudinal perspective lies in three fundamental contributions to the HCI literature:

- The Legitimization Shift: We document the transition of surveillance from a contested social practice (2022) to a regulated household standard (2025). Narratives of efficiency and convenience have been replaced by a logic of 'household data management,' effectively removing surveillance from the realm of social negotiation.
- The Ambient Erasure of Agency: The 'delta' between our two study phases shows the shift from discrete devices to infrastructure-level sensing (smart locks/lights). This has removed the physical 'opt-out' mechanisms available in 2022, leading workers toward performative adaptation rather than genuine privacy.
- The Governance Gap: We theorize the 'Responsibility Shift,' where the 2023 Law offloads the burden of governance onto the household. Within these spaces, legal principles are consistently superseded by the Kafala system's existing power structures, weaponizing transparency to enforce control.

Ultimately, this paper proves that in high-asymmetry environments, transparency-based regulation acts as a catalyst for power formalization. The 'Compliance Paradox' serves as a critical warning: without addressing underlying labour inequalities, digital rights frameworks may inadvertently provide the 'veneer of legitimacy' needed to deepen domestic intrusion.

## 6. DISCUSSION

This study contributes to understanding how smart home technologies reshape privacy, surveillance, and power dynamics within domestic environments. Drawing on a longitudinal qualitative analysis of Jordanian households, we identify the compliance paradox: the introduction of data protection regulation increases awareness of privacy while simultaneously enabling new forms of surveillance through ambiguity, selective interpretation, socio-cultural mediation, and evolving technological affordances. Rather than resolving tensions between privacy and control, regulation reconfigures how these tensions are expressed, justified, and managed in everyday life.

More broadly, our findings suggest that privacy governance in smart homes cannot be understood as a matter of legal compliance or technical safeguards alone. Instead, governance emerges through the interaction of regulation, household authority, technological infrastructures, and everyday social practices. This shifts attention from privacy as an individual right to privacy as a socio-technical condition structured by power.

### 6.1 Reconfiguring Surveillance Through Regulation

A central contribution of this study is showing that regulation does not simply constrain surveillance practices but can actively reshape and legitimize them. Contrary to assumptions in privacy research that legal frameworks inherently strengthen user protection, our findings show that regulation may provide discursive resources for justifying monitoring practices.

Participants often interpreted the absence of explicit restrictions as implicit permission to engage in surveillance. This marks a shift from moral and security-based justifications identified in a prior study toward legality-based reasoning, where compliance is constructed through interpretation rather than enforcement. Rather than challenging surveillance, regulation becomes a flexible framework strategically mobilized to support existing practices.

This finding complicates assumptions that stronger regulation necessarily produces stronger protection. Under conditions of ambiguity, regulation may generate new legitimacy even when it fails to constrain harmful practices. Our findings suggest regulatory ambiguity is not merely a gap in governance, but a productive condition shaping how power operates through selective interpretation and the strategic mobilization of legality. This helps explain why the compliance paradox is not simply a failure of enforcement, but a structural feature of socio-technical governance under ambiguity.

### 6.2 From Episodic Monitoring to Ambient Datafication

Our findings show that the compliance paradox is inseparable from transformations in data collection itself. Building on the episodic and device-specific monitoring identified in a prior study, we find that between 2022 and 2025, monitoring shifted toward ambient and persistent data collection through interconnected smart home ecosystems. This transforms domestic environments into data-rich spaces where activities are continuously recorded, inferred, and linked across devices.

This shift alters not only the scale of surveillance, but its ontology. Surveillance is no longer experienced simply as visible observation, but as an infrastructural condition embedded in everyday life. Rather than discrete acts tied to specific devices or moments, ambient datafication distributes surveillance across interconnected systems operating continuously, often passively and opaquely. Data collection increasingly enables interpretation, profiling, and evaluation, extending surveillance beyond what people do to what their actions are taken to mean.

These findings also extend the compliance paradox. Regulation may appear to govern surveillance, yet transformations in data collection can expand monitoring faster than legal frameworks can address. Law may regulate visible surveillance while leaving ambient and inferential forms of data power comparatively unchallenged.

### 6.3 Domestic Spaces as Sites of Informal Governance

Our findings highlight domestic environments as sites of informal governance, where technological practices are shaped by socio-cultural norms rather than formal institutional mechanisms. In the Jordanian context, this informal governance is inextricably linked to the Kafala (Sponsorship) system, which creates a structural, state-sanctioned dependency between the employer (sponsor) and the domestic worker.

Unlike organizational settings, where surveillance is mediated through policies and oversight structures, smart homes operate through interpersonal relationships, implicit expectations, and culturally situated authority structures. Governance does not disappear in domestic spaces—it becomes privatized. This insight helps explain why formal regulation, such as Jordan's 2023 Data Protection Law, has limited impact. Legal rules do not operate directly, but are interpreted and filtered through household authority.

Our findings therefore suggest domestic spaces should be understood not as zones outside governance, but as governance sites in their own right. This contributes to HCI and governance scholarship by suggesting privacy governance should be understood not only as institutional regulation, but as an everyday socio-cultural practice shaped through informal, relational, and culturally embedded mechanisms.

### 6.4 From Observation to Internalized Control

The longitudinal comparison reveals a transformation in how surveillance is experienced by domestic workers. In a prior study, workers expressed discomfort and occasional resistance. In the follow-up study, resistance becomes less visible, replaced by adaptation, internalization, and self-regulation. Workers increasingly modify behaviour in anticipation of being observed, showing how surveillance operates through internalized compliance rather than direct enforcement alone.

Adaptation should not be confused with acceptance, nor internalization with genuine consent. What appears as compliance often reflects constrained responses to unequal conditions, where resistance carries social or economic risks. This shift suggests surveillance increasingly operates through anticipation. Workers act not only in response to being watched, but in relation to the possibility of being watched. This embeds control in ordinary routines, self-monitoring, and behavioural adjustment.

This also complicates conventional understandings of agency. Rather than framing subjects as either resisting or complying, our findings suggest adaptation may represent constrained negotiation under unequal conditions, particularly where economic dependency and social vulnerability (exacerbated by Kafala structures) shape possible responses.

### 6.5 Socio-Digital Inequality and Behavioural Profiling

A further contribution of this study is showing that privacy vulnerability is structured by socio-digital inequality. Differences in technological literacy, legal awareness, and access to resources shape who can understand, contest, or benefit from smart home systems. Privacy is not experienced symmetrically across occupants of the same smart home. Those with greater technological knowledge or control over devices often possess disproportionate authority over data practices.

At the same time, our findings show surveillance increasingly operates through behavioural profiling, where ordinary actions are transformed into judgments about trust, discipline, or performance. This extends workplace surveillance logics into domestic environments, but without protections associated with formal employment. Profiling and socio-digital inequality also reinforce one another. Those with less capacity to understand or contest data practices are often most vulnerable to classification and judgment through opaque monitoring systems, creating a compounding dynamic in which inequality and surveillance power intensify.

### 6.6 Implications for Design

These findings suggest moving beyond consent-centered approaches to smart home privacy design. Existing approaches often assume individual users making informed choices about data collection, yet smart homes involve multiple occupants with unequal power, divergent interests, and differing capacities to understand or contest surveillance. Privacy design therefore requires relational and context-sensitive approaches.

First, designers should support multi-user privacy, rethinking access control, permissions, and privacy settings to account for co-present users whose interests may conflict. Second, designers should make ambient data collection visible through interfaces that expose what is collected, where data flows, who has access, and how inferences or profiling judgments are produced. Third, systems should support negotiation and contestability, including mechanisms that allow vulnerable occupants to raise concerns, express preferences, or access shared controls.

Fourth, designers should consider safeguards against behavioural profiling, including minimizing unnecessary inference, limiting evaluative uses of sensor data, and making profiling logics transparent. Fifth, design should explicitly account for socio-digital inequality. Sixth, our findings suggest designing for power-aware privacy. This includes a shift toward frictional agency and physical overrides—such as mechanical lens covers—that provide vulnerable occupants with 'micro-zones' of privacy that are mechanically, rather than just legally, guaranteed.

Finally, these findings suggest a broader shift from designing for compliance toward designing for accountability. Because regulation may be ambiguous or weakly enforced, design cannot assume legal protections alone will mitigate harms.

### 6.7 Implications for Policy

These findings suggest privacy regulation should move beyond organizational models that overlook domestic environments. Existing data protection frameworks often assume harms arise primarily through commercial or institutional practices, yet our findings show significant privacy harms can emerge within domestic settings shaped by informal governance, unequal power relations, and weak oversight.

First, laws should explicitly address household surveillance and clarify how data protection principles apply within smart homes. Second, stronger enforcement mechanisms are needed, including reporting channels, oversight mechanisms, and remedies accessible to vulnerable populations. Third, policy should address power asymmetries directly, recognizing consent and autonomy cannot be assumed in contexts of dependency.

Fourth, policymakers should address behavioural profiling and ambient data collection explicitly, rather than relying on generic privacy principles poorly suited to harms arising through inference and continuous monitoring. Fifth, policy must address not only whether rules exist, but how they are interpreted, enacted, and appropriated in practice. Sixth, policy should recognize domestic workers and similarly vulnerable occupants not merely as incidental subjects, but as stakeholders requiring explicit protection, potentially through closer integration of labour protections, privacy protections, and digital rights.

More broadly, these findings suggest moving from abstract rights toward context-sensitive governance attentive to how law interacts with culture, informal authority, technological infrastructures, and everyday social practices.

### 6.8 Implications for Users and Social Practices

Our findings also have implications for users and everyday social practices. While privacy is often framed as a matter of legal rights or technical controls, our findings suggest privacy in smart homes is also shaped through ordinary social practices: how monitoring is normalized, how authority is exercised, how technologies are interpreted, and how occupants negotiate shared domestic life.

For households, these findings challenge assumptions that surveillance is simply a neutral tool for convenience or security. Monitoring practices are relational practices with consequences for trust, autonomy, dignity, and power. For domestic workers and vulnerable occupants, the findings underscore the importance of privacy awareness, legal knowledge, and collective support structures.

More broadly, these findings suggest privacy harms are reproduced not only through technologies or weak laws, but through social practices that make surveillance appear ordinary or unavoidable. Addressing surveillance harms may therefore require cultural as well as legal and technical change.

## 7. CONCLUSION

This study examined how smart home technologies reshape privacy, surveillance, and power dynamics within Jordanian households through a longitudinal analysis spanning the period before and after Jordan's 2023 Data Protection Law. Comparing data from 2022 and 2025, we demonstrate that regulatory and technological changes do not simply mitigate surveillance harms; they fundamentally transform the discursive and structural logic through which those harms are justified and enacted.

A central contribution of this work is the identification of the compliance paradox: the phenomenon where data protection regulation increases privacy awareness while simultaneously enabling new forms of surveillance. We show that regulation is not a neutral constraint but a strategic resource—a 'legalistic veneer'—that actors mobilize to normalize monitoring. This paradox is especially acute when legal frameworks are filtered through the structural dependencies of the Kafala (Sponsorship) system, where formal privacy rights are consistently superseded by labour-structural power.

Our findings further reveal a qualitative shift in the domestic ontology from episodic monitoring to ambient datafication. As surveillance becomes embedded in the home's infrastructure, resistance becomes less visible, replaced by the internalized self-regulation of vulnerable occupants. This transition demonstrates that power increasingly operates through the authority to define conduct via opaque behavioural profiling and digital inference. These results suggest a critical need for smart home designs attentive to multi-user privacy and contestability, alongside policy frameworks that explicitly address the governance gap in domestic labour.

While our findings are grounded in the specific socio-cultural and regulatory context of Jordan, this case serves as a prognostic laboratory for global privacy governance. It offers vital insights for any context where labour precarity and residency status are linked—including seasonal worker programs or visa-dependent employment in Western nations. Future research should investigate how compliance paradoxes emerge in other smart environments and explore design interventions, such as 'frictional agency,' that are truly responsive to relational vulnerability and informal authority.

Ultimately, as domestic environments become increasingly datafied, the central challenge is not simply protecting data through stronger compliance. It is confronting how surveillance, regulation, and everyday power have become intertwined in ways that reproduce, rather than reduce, inequality.

## 8. DECLARATION OF GENERATIVE AI AND AI-ASSISTED TECHNOLOGIES

During the preparation of this manuscript, the authors used ChatGPT and Gemini to assist with proofreading and language refinement. All content generated with this tool was subsequently reviewed and revised by the author(s), who take full responsibility for the final version of the published article.